\documentclass{article}

\PassOptionsToPackage{numbers, compress}{natbib}
\usepackage[final,sglblindworkshop]{neurips_2026}
\usepackage[utf8]{inputenc}
\usepackage[T1]{fontenc}
\usepackage{microtype}
\usepackage{graphicx}
\usepackage{subcaption}
\usepackage{booktabs}
\usepackage{multirow}
\usepackage{placeins}
\usepackage{amsmath}
\usepackage{amssymb}
\usepackage{mathtools}
\usepackage{url}
\usepackage{hyperref}
\usepackage{todonotes}

\title{Calibrating AlphaFold confidence scores \\for interface quality}

\workshoptitle{ML4Molecules}

\author{%
\textbf{Sagyndyk Sagadinov}\,$^{1}$\quad
Vsevolod Viliuga
\thanks{Corresponding author. Contact: vsevolod.viliuga@scilifelab.se}\,\,\,$^{1,2}$
\quad
\textbf{Arne Elofsson}\,$^{1}$ \\
\\$^1$ DBB and SciLifeLab at Stockholm University, Stockholm, Sweden\\
$^2$Max Planck Institute for Polymer Research, Mainz, Germany}

\begin{document}

\maketitle
\vspace{-0.2cm}

\begin{abstract}

AlphaFold confidence scores are widely used to rank protein complex predictions and to select subsets of models for downstream applications. Although these scores provide strong discriminatory power for distinguishing prediction quality, their numerical values were never intended to represent the probability that a prediction satisfies a given structural quality criterion. In this work, we calibrate commonly used confidence scores to estimate the probability that a predicted complex exceeds a chosen threshold of DockQ, an established measure of interface quality. On antibody-antigen and general multimer datasets, post-hoc isotonic calibration consistently improves agreement between predicted probabilities and observed DockQ success rates while preserving the underlying score ordering. The calibrated probabilities also provide more accurate estimates of the fraction of selected models expected to pass a given structural quality threshold. Although the mapping is fitted within a given prediction distribution and therefore transfers imperfectly across prediction domains, we show that reliable estimates are recovered from a small number of labeled target-domain samples. Because the calibration is entirely post hoc and requires no underlying AlphaFold model fine-tuning, it provides a straightforward and customizable way to convert existing model quality rankings into calibrated probabilities of passing a chosen structural quality threshold.

% Even though calibration might be sensitive to domain shift, we demonstrate that reliable probability estimates can be recovered from a relatively small amount of labeled samples representative of the target prediction domain. 

\end{abstract}

\section{Introduction}

Recent protein structure prediction models, including the AlphaFold family \citep{jumper2021highly,evans2021protein,abramson2024accurate} and subsequent related approaches \citep{wohlwend2025boltz,chai2024chai,candido2026language}, provide confidence scores that have become a gold standard for ranking and selecting predicted structures. These scores are particularly important for protein complexes, where the quality
of the predicted interaction can substantially vary even if the individual protein chains are modeled correctly. Interface-focused confidence measures such as ipTM and its derivatives (e.g. ipSAE \citep{dunbrack2025ipsae} or actifpTM \citep{varga2025actifptm}), are therefore commonly used to prioritize complexes that are more likely to contain a correctly modeled interface.

One of the most established ways to assess the actual quality of a predicted protein-protein interface against a ground truth complex is the DockQ score \citep{mirabello2024dockq}. Importantly, AlphaFold confidence scores were never trained to reproduce it: ipTM estimates an interface TM-score \citep{zhang2004scoring}, a continuous structural similarity rather than the probability of any event. Its values nonetheless fall $\in [0, 1]$ and are widely used as selection cutoffs, where a high score is often interpreted to indicate a correspondingly high chance that the selected model is correct. This motivated the development of direct DockQ prediction approaches based on confidence features, such as pDockQ \cite{bryant2022improved} and pDockQ2 \citep{zhu2023evaluation}. However, neither these predicted quality nor the original confidence scores directly provide a calibrated probability that a model exceeds a specific DockQ threshold. Consequently, applying a fixed score cutoff, as commonly done in practice \cite{yin2022benchmarking, o2023protein, sasvari2024neutrophil, han2026alphafold}, does not quantify how many selected models are expected to satisfy the desired structural quality criterion.

In this work, we use post-hoc calibration to assign explicit probabilistic interpretation to confidence scores. A score that ranks well but carries no probabilistic meaning is an established setting for post-hoc calibration \citep{platt1999probabilistic, niculescu2005predicting, guo2017calibration}. Given a score \(s\) and labeled samples representative of the target prediction domain, we fit a threshold-specific mapping 
\[
g_t(s)=\hat{P}\!\left(\mathrm{DockQ}\ge t \mid s\right).
\]
To do so, we use simple isotonic regression \citep{zadrozny2002transforming}, which preserves the original score ordering up to ties. The calibrated score can then be interpreted as the estimated probability that a prediction exceeds the chosen DockQ threshold.

\paragraph{Main contributions}
We first show that confidence scores, which discriminate model quality well, can still be far from the observed success frequencies. Isotonic calibration, however, turns them into accurate probabilities of passing a certain interface quality threshold. We further show that this gap has a direct practical cost: for models selected by a confidence cutoff or by top-$K$ ranking, reading the ranking score values as success probabilities substantially mis-estimates how many of the selected models will actually pass. Under distribution shift, the fitted mapping may transfer poorly, but we find that reliable threshold probabilities can be recovered from a limited labeled sample set drawn from the target distribution. Because calibration is entirely post hoc, it requires no expensive model fine-tuning and provides a straightforward and customizable way to adapt existing confidence scores to a specific application domain.

\section{Calibration and evaluation settings}
\label{sec:calibration_settings}

\paragraph{Calibration datasets}
The Ab-Ag dataset \citep{fromm2026evaluating} contains AlphaFold3 predictions for antibody-antigen complexes split temporally into train/test from the SAbDab database\citep{dunbar2014sabdab}. Its held-out test set contains 62,681 models for 1,257 targets. The general multimers dataset contains 5,496 AlphaFold-Multimer predictions from 1,922 targets and is evaluated using target-grouped five out-of-fold predictions \citep{zhu2023evaluation}. For external evaluation, we use the dataset published in \citet{han2026alphafold} where AFDB entries with experimentally validated structures were used as positive examples for benchmarking confidence scores against structural ground truth. We retain the 304 experimentally supported complexes with successfully computed DockQ values, which constitute the AFDB benchmark used here. Further details on dataset construction are provided in App.~\ref{app:benchmark_details}. 
%\todo{explanation how the AFDB dataset was prepared in appendix please}

\paragraph{DockQ interface quality criteria}
Interface quality is evaluated using DockQ v2 \citep{mirabello2024dockq}. We consider the standard CAPRI thresholds \citep{lensink2023capri}: DockQ $\geq0.23$, $\geq0.49$, and $\geq0.80$, corresponding to acceptable, medium, and high structural quality. For Ab-Ag complexes, antibody chains are treated as one interaction partner and antigen chains as the other for the interface DockQ computation.

\paragraph{Post-hoc calibration}
For each DockQ threshold $t \in \{0.23, 0.49, 0.80\}$, isotonic regression fits a monotone mapping
\begin{equation}
    \hat p_i^{(t)} = g_t(s_i),
\end{equation}
where $\hat p_i^{(t)}$ estimates $P(\mathrm{DockQ}_i\geq t\mid s_i)$ from the underlying confidence score $s_i$. We provide more details in App.~\ref{app:calibration_details}.

\paragraph{Confidence-score baselines and metrics}
We compare ipTM with published pDockQ2 \citep{zhu2023evaluation}, ipSAEmax \citep{dunbrack2025ipsae}, and an Ab-Ag-specific pDockQ2 refit (details in App. \ref{app:pdockq2_refit}). We provide extensive details on confidence scores definitions and their usage in App. \ref{app:score_definitions}. Probability calibration is evaluated using Brier score and expected calibration error (ECE) (definitions in App. \ref{app:calibration_details}).
%Additional baseline and classification metrics are reported in the appendix.

\paragraph{Success rate estimation after model selection}
We evaluate whether confidence scores can estimate the fraction of selected models that pass a chosen DockQ threshold. For any selected set $S$, the observed success fraction is
\begin{equation}
    Y^{(t)}(S)= \frac{1}{|S|} \sum_{i\in S} \mathbf{1}[\mathrm{DockQ}_i\geq t],
\end{equation}
and the (un)-calibrated estimate is
\begin{equation}
    \hat Y^{(t)}(S)=\frac{1}{|S|}\sum_{i\in S}\hat p_i^{(t)}.
\end{equation}

% \textbf{Score-cutoff selection (Fig.~\ref{fig:reliability}).}
% For each raw ipTM cutoff $c$, we select all models with $\mathrm{ipTM}\geq c$. This set is defined once and kept fixed across estimators. We then compare the observed fraction passing each DockQ threshold with the fraction estimated from raw ipTM, pDockQ2, and threshold-specific isotonic probabilities as the selection cutoff is varied.

% \textbf{Top-$K$ selection (Fig.~\ref{fig:fixedset}).}
% We separately consider the common setting in which a fixed number of highest-ranked models is retained for each target. Models are ranked by raw ipTM, and the top-$K$ set is again kept fixed across estimators. We evaluate $K\in\{5,10,20,50\}$ for Ab--Ag and $K=5$ for general multimers, comparing the observed and estimated fractions that pass each DockQ threshold.
\paragraph{AFDB transfer and local recalibration}
To evaluate calibration under distribution shift, we freeze the general multimers-fitted isotonic ipTM mapping and apply it directly to AFDB without refitting. We then recalibrate on progressively larger labeled subsets of AFDB and evaluate each fitted mapping on the remaining held-out complexes. The same splits are used to calibrate ipSAEmax. Full protocol details are provided in App.~\ref{app:calibration_details}.
\section{Results}

% \subsection{Confidence scores do not directly quantify the probability of passing a structural quality threshold}

\subsection{Isotonic calibration maps confidence scores to probabilities of passing structural quality thresholds}

\begin{figure}
\centering
\includegraphics[width=1.0\textwidth]{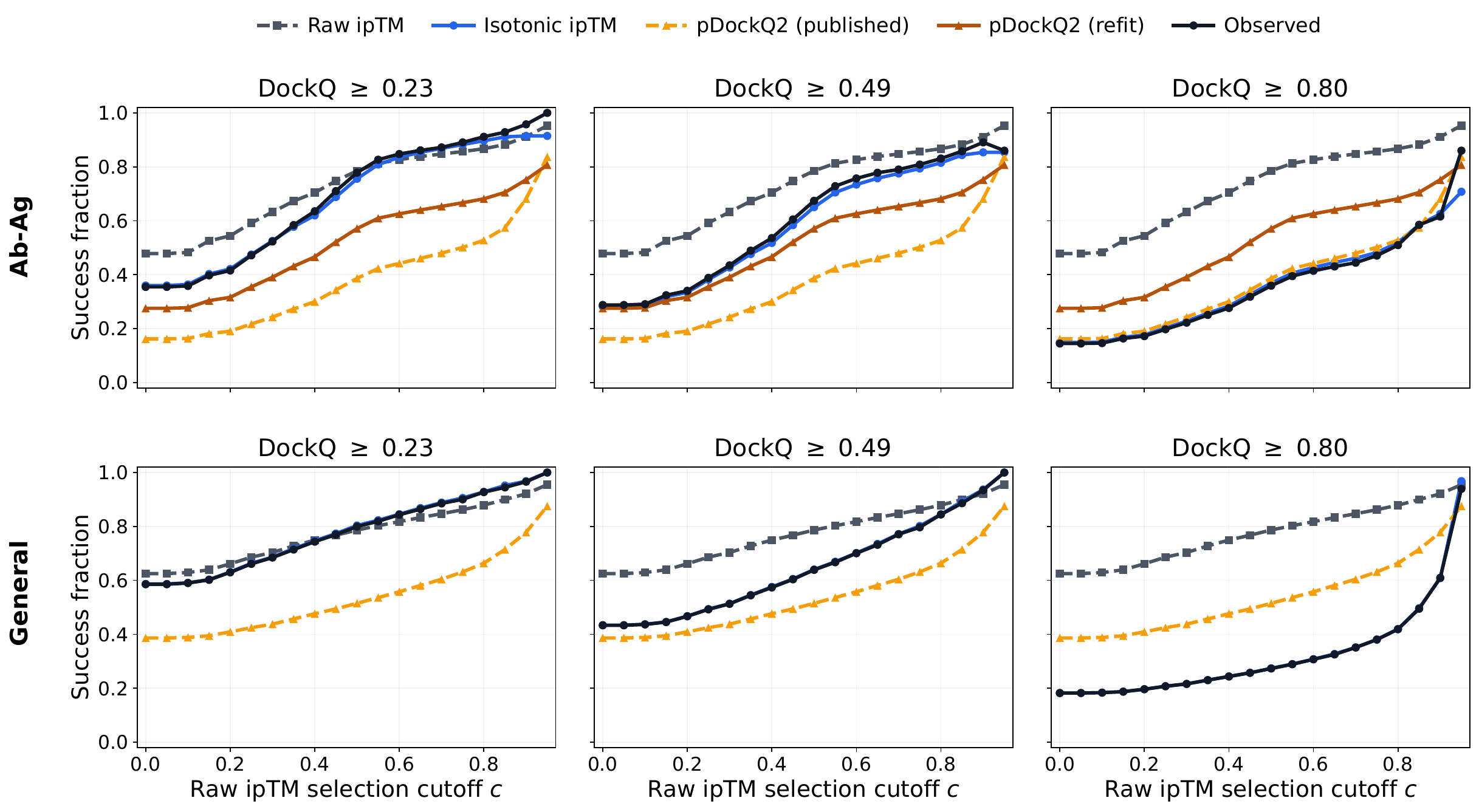}
\caption{Fraction of selected models passing each DockQ threshold as a function of the raw ipTM selection cutoff. For each cutoff, all models with ipTM above the threshold are selected and kept fixed. Isotonic ipTM closely follows the DockQ threshold-dependent real success rate, whereas direct interpretations of raw ipTM and pDockQ2 as probabilities can substantially over- or underestimate it.}
\label{fig:reliability}
\end{figure}

% Although the evaluated confidence and ranking scores succeed at accurately discriminating the model quality, their numerical values do not necessarily reflect the probability of passing a structural-quality threshold, indicated by high AUROC but also high calibration errors in Tab.~\ref{tab:abag_baselines}. Fig.~\ref{fig:reliability} illustrates this distinction using ipTM, by treating the raw score as a naive estimate of the probability of passing each DockQ threshold and comparing it with the observed success frequency. Despite strong discrimination, directly interpreting raw ipTM as a probability proxy leads to substantial disagreement with the observed success frequency, particularly for stricter DockQ criteria. Under this naive interpretation, for example, an ipTM value of 0.7 would correspond to a 70\% probability of passing the chosen DockQ threshold, which is not supported by the empirical frequencies. Isotonic calibration instead learns a threshold-specific mapping from ipTM to \(P(\mathrm{DockQ}\ge t)\), while preserving the original ordering (Tab.~\ref{tab:abag_baselines}, AUROC column).

Although the evaluated confidence and ranking scores discriminate model quality well, with high AUROC throughout, the calibration errors are substantial if their values are read directly as success probabilities (Tab.~\ref{tab:abag_baselines}, \ref{tab:general_baselines}).
To demonstrate its practical consequence, we selected all models exceeding a series of ipTM cutoffs on both Ab-Ag and general multimer datasets and reported the fraction of samples that actually passed each DockQ criterion. For the same fixed ipTM cutoff, the real (observed) success fraction changes substantially with the chosen DockQ threshold (Fig.~\ref{fig:reliability}). Because the selected set is identical across DockQ criteria, a single ipTM value cannot express this threshold dependence. Reading it as a success probability therefore overestimates the number of successful models, particularly for stricter
DockQ criteria. The same holds for scores that were fitted to predict DockQ where pDockQ2 over- or underestimates the observed success fraction depending on the threshold (Fig.~\ref{fig:reliability}), because a single continuous quality prediction cannot encode $P(\mathrm{DockQ} \ge t)$ for several $t$ simultaneously. In contrast, threshold-specific isotonic calibration directly maps the underlying score to $P(\mathrm{DockQ}\geq t)$, and the resulting calibrated estimates closely follow the observed success fraction across the same model selection while preserving the original ordering up to ties (see App. \ref{sec:ranking_preservation}).

\begin{figure}[t]
\centering
\includegraphics[width=1.0\textwidth]{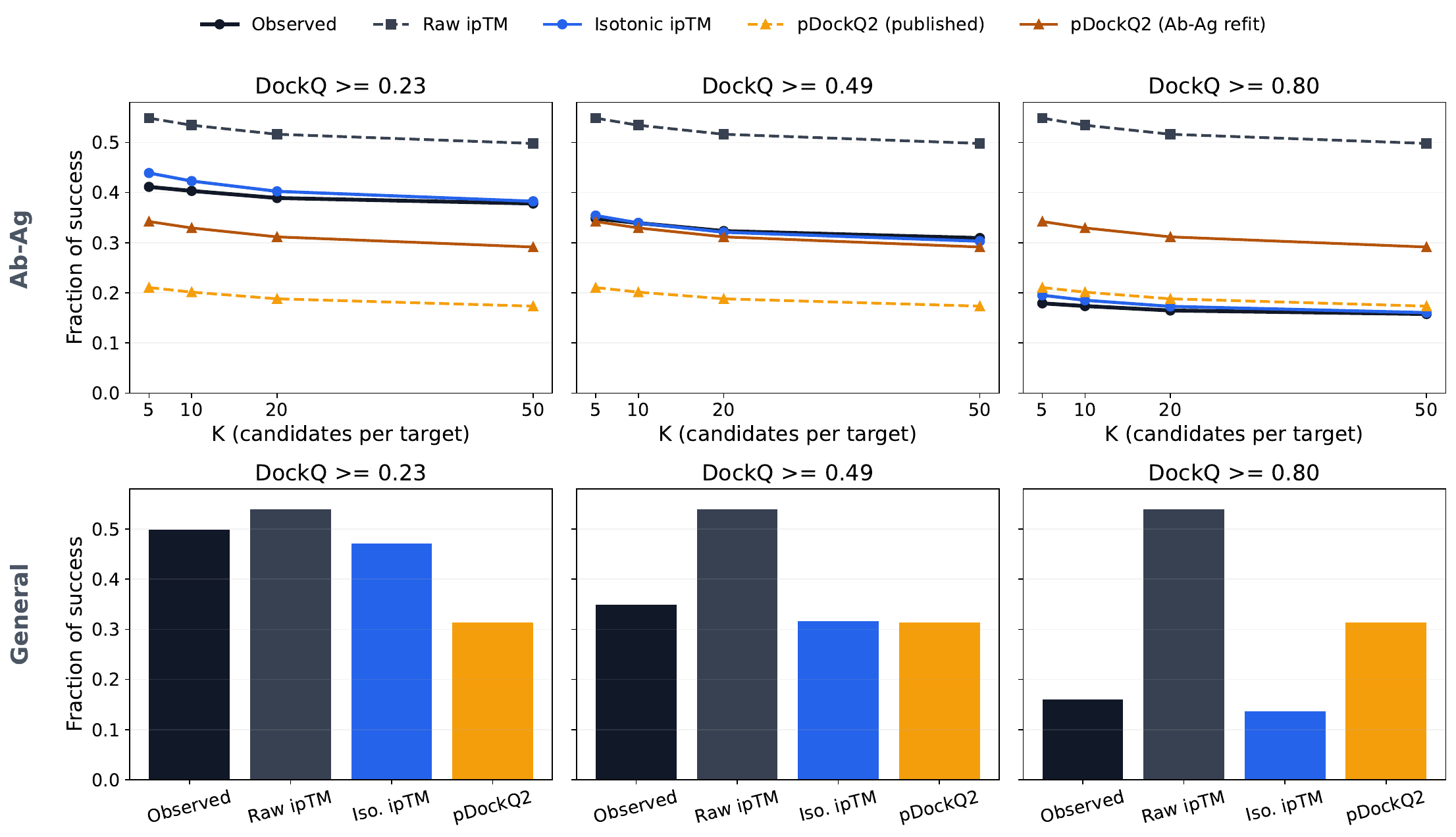}
\caption{Fraction of selected models passing each DockQ threshold, analogous to Fig. \ref{fig:reliability}, but as a function of the top-$K$ candidates ranked by ipTM. Closer to observed is better.}
\label{fig:fixedset}
\end{figure}

% In another commonly used setting, we selected the top 50 models per Ab--Ag target and the top 5 models per General multimer target and asked how many of the selected predictions are expected to pass each DockQ threshold (Fig.~\ref{fig:fixedset}). If the score values are interpreted directly as probabilities, their average gives the expected success fraction and their sum gives the expected number of successful models. This substantially misestimates the fraction that actually passes the DockQ criterion for raw ipTM and, depending on the threshold, for pDockQ2. In contrast, summing the threshold-specific isotonic probabilities closely recovers the observed number of successful models in the selected sets.

Another common setting in protein structure prediction is to select the top-$K$ models by the confidence score for downstream analysis or experimental validation. We therefore computed the expected success fraction within the top-$K$ models selected by ipTM, using up to $K=50$ for Ab-Ag and $K=5$ (maximum number of samples available) for general multimers. If the score values are interpreted directly as probabilities, averaging over them gives the expected success fraction. As expected, this substantially mis-estimates the fraction of models that passes the DockQ criterion for raw ipTM and, depending on the threshold, for pDockQ2 (Fig.~\ref{fig:fixedset}). In contrast, the threshold-specific isotonic probabilities closely recover the observed success fraction also among top-$K$ models.

\subsection{Transfer and recalibration across prediction domains}

\label{sec:local_calibration}

\begin{figure}[t]
\centering
\includegraphics[width=1.0\textwidth]{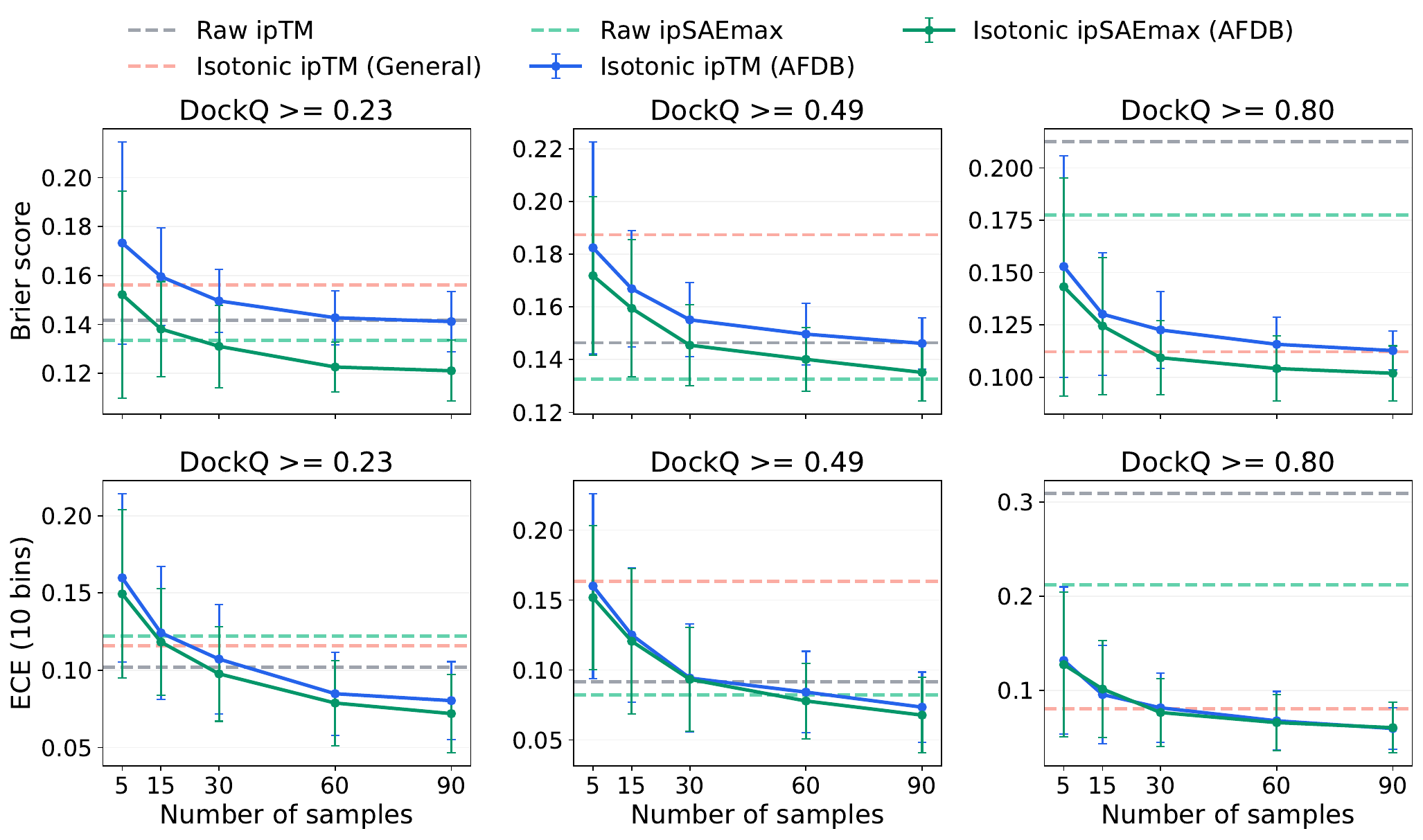}
\caption{Recalibration on limited number of samples from AFDB. Brier score (top) and ECE (bottom) are shown for DockQ thresholds of 0.23, 0.49, and 0.80 as a function of the number of labeled AFDB examples used for isotonic calibration. Curves show mean $\pm$ standard deviation over 100 stratified splits. Reference lines show the corresponding raw score and frozen general-to-AFDB performance.}
\label{fig:afdb_recalibration}
\end{figure}

\paragraph{Transfer of isotonic calibration beyond the training distribution is limited}
Since isotonic calibration estimates the relationship between a confidence score and the probability of passing a given DockQ threshold within a given distribution of predictions, the fitted mapping may change under distribution shift. To test its transferability, we fitted the isotonic ipTM mapping on the general multimer dataset, froze it, and applied it directly to the external AFDB dataset. We find that the transferred mapping, in general, no longer reproduces the observed DockQ success frequencies consistently across quality thresholds on the distribution of AFDB models (Table~\ref{tab:afdb}). At DockQ $\geq0.23$ and $\geq0.49$, the transferred mapping is even a worse probability proxy compared with the raw ipTM, while at DockQ $\geq0.80$ it remains more closely aligned with the observed success. 

\paragraph{Recalibration with limited target-domain data}

% OLD:
% We next tested how much labeled data from the target distribution is required to recover a reliable confidence score-probability mapping. On the AFDB dataset, we fitted ipTM using progressively larger fractions of labeled target domain samples and evaluated each fitted mapping on the remaining held-out complexes (Fig.~\ref{fig:afdb_recalibration}). Calibration improves rapidly as more samples are used, with most of the gain occurring within the first $20\%$ of labeled samples, corresponding to approximately 60 samples.Since AFDB contains only about 300 complexes, this corresponds to roughly 60 labeled structures. Because isotonic calibration can be applied to any confidence score, we also calibrated ipSAEmax\todo{section with definition}, which provides the strongest discrimination on AFDB among the evaluated scores, and observed the same trend (Fig.~\ref{fig:afdb_recalibration}, ipSAEmax). Because isotonic calibration can be applied to any confidence score, we also calibrated ipSAEmax\todo{section with definition}, which provides the strongest discrimination on AFDB among the evaluated scores, and observed the same trend (Fig.~\ref{fig:afdb_recalibration}, ipSAEmax).

% NEW:
We next tested how much labeled data from the target distribution is required to recover a reliable confidence score-probability mapping. On AFDB, which contains 304 complexes with computed DockQ values, we fitted isotonic mappings on calibration sets of 5 to 90 complexes and evaluated each on the remaining held-out complexes. The amount of labeled data required depends on how poorly the raw score is calibrated to begin with. At the high-quality criterion, where raw ipTM departs furthest from the observed success frequencies, local recalibration is already substantially better than both raw references at the smallest calibration sets we evaluate, and both Brier score and ECE continue to improve as more samples are added. At the acceptable and medium thresholds, the raw scores are comparatively well calibrated on AFDB, and local recalibration improves over them in ECE at around 30 samples or approximately $10\%$ of the available complexes. Brier score improves more slowly. Because isotonic calibration can be applied to any confidence score, we also calibrated ipSAEmax, which provides the strongest discrimination on AFDB among the evaluated scores (Tab. \ref{tab:afdb}). It follows the same trend and gives the lower Brier score of the two calibrated signals at the acceptable and high-quality criteria (Fig.~\ref{fig:afdb_recalibration}).

The same behavior is observed on the much larger Ab-Ag dataset. Isotonic ipTM improves over raw ipTM in both Brier score and ECE at every threshold using only 50 labeled models, and performance largely saturates at roughly 500--1000 models (Fig.~\ref{fig:abag_recalibration}). Increasing the calibration set to 10K models, still under 5\% of the available training data, gives almost no further benefit. Since models predicted for the same target are highly correlated, the effective sample size is closer to the number of unique protein complexes than to the number of models. Together, these results indicate that reliable score-probability mappings can be fitted from a relatively small number of representative samples drawn from the target distribution and that the largest gains occur where the raw score departs most from the observed success frequencies.

\section{Discussion and Conclusion}

Confidence scores for protein-complex prediction are widely used for ranking and filtering models, yet their numerical values do not directly represent the probability that a prediction passes a certain structural quality check. We show that post-hoc isotonic calibration maps these scores to threshold-specific estimates of $P(\mathrm{DockQ}\geq t)$ while preserving their ordering (up to ties). This gives the scores a direct operational interpretation: for models selected by a confidence score cutoff or by top-$K$ ranking, calibrated probabilities closely estimate the fraction or number of predictions expected to pass the chosen DockQ threshold.

The main limitation is sensitivity to distribution shift since a score-probability mapping fitted in one domain may no longer reproduce the observed success frequencies in another. Our recalibration experiments indicate that this limitation can be addressed with relatively few labeled samples approximating the target-domain distribution. On AFDB, calibration improves rapidly using only a fraction of the roughly 300 available complexes, while on the much larger Ab--Ag dataset the performance largely saturates and on Ab--Ag calibration within a few hundred labeled models. This suggests that accurate calibration does not require large training datasets or expensive fine-tuning of the underlying structure-prediction model. Because the procedure is post-hoc and score-agnostic, it can be applied to whichever confidence signal turns out to be most informative within the target domain, as we demonstrate for ipSAEmax. Within the scope of structural quality defined by DockQ, calibration therefore provides an easy and customizable way to connect confidence-based model selection with an explicit probabilistic estimate of how many selected predictions are expected to satisfy the desired interface quality criterion.

% Bibliography
\bibliographystyle{unsrtnat} % plainnat does not cite misc/arxiv articles properly
\bibliography{references}

\appendix
\clearpage
\appendix

\section{Details on dataset construction}
% \section{Ben-}
\label{app:benchmark_details}
 
\paragraph{Ab-Ag}
The Ab--Ag dataset is derived from SAbDab \cite{dunbar2014sabdab} and follows the temporal split of \citet{fromm2026evaluating}. The training chunk used for calibration contains 269,105 AlphaFold3 model predictions from 7045 unique antibody-antigen targets. The held-out temporal test set contains 62,681 model predictions from 1,257 targets. All mappings evaluated on the held-out Ab-Ag test set are fitted using only the training split.

\paragraph{General multimers}
The General multimers dataset is derived from the PDB-based AlphaFold-Multimer benchmark of \citet{zhu2023evaluation}. The original benchmark was constructed from experimentally determined complexes released after the AlphaFold training cutoff, followed by redundancy and homology reduction against the training set. Its final modeled dataset contains 1928 complexes with two to six chains: 1148 dimers, 220 trimers, 367 tetramers, 62 pentamers, and 131 hexamers, including both homomeric and heteromeric assemblies. The processed dataset used in our calibration analysis contains 1922 complex targets represented by 5496 chain-specific structural quality observations. We use unique complex identity for grouping of the five-fold out-of-fold calibration, such that all observations originating from the same complex are assigned to the same fold.

\paragraph{External AFDB dataset}
The confidence benchmark of \citet{han2026alphafold} contains 691 predicted dimers: 230 experimentally supported homodimers paired with 117 monomeric controls, and 94 experimentally supported heterodimers paired with 250 non-interacting chain-pair controls. Thus, 324 entries correspond to experimentally supported complex structures. Because our analysis evaluates structural quality using DockQ against an experimentally determined complex structure, the monomeric and non-interacting controls are not included in the structural-quality evaluation. DockQ was successfully computed for 304 of the 324 experimentally supported complexes, the remaining 20 entries where DockQ computation failed were excluded. These 304 complexes constitute the AFDB structural-quality benchmark used in this work.

% \paragraph{AFDB dataset construction}
% The confidence benchmark of \citet{han2026alphafold} contains 691 predicted dimers: 230 experimentally supported homodimers paired with 117 monomeric controls, and 94 experimentally supported heterodimers paired with 250 non-interacting chain-pair controls. Thus, 324 entries correspond to experimentally supported complex structures. Because our analysis evaluates structural quality using DockQ against an experimentally determined complex structure, the monomeric and non-interacting controls are not included in the evaluation. DockQ was successfully computed for 304 of the 324 complexes. These 304 complexes constitute the AFDB structural quality benchmark used in this work.

\section{Details on confidence scores}
\label{app:score_definitions}

\paragraph{ipTM.}
The interface predicted TM-score (ipTM) is an AlphaFold confidence score for the relative arrangement of chains. It applies a TM-score-like transformation to cross-chain predicted aligned errors (PAE) and aggregates these values across residues:
\[
\mathrm{ipTM}
=
\max_i
\frac{1}{|D_i|}
\sum_{j\in D_i}
\frac{1}{
1+\left(\mathrm{PAE}_{ij}/d_0(|D_i|)\right)^2
},
\]
where $D_i$ contains residues belonging to chains other than that of residue $i$, and $d_0$ is the usual length-dependent TM-score normalization. Thus, ipTM summarizes confidence in the inter-chain arrangement, but its numerical value is not fitted to the probability of satisfying a DockQ criterion \citep{evans2021protein}.

\paragraph{pDockQ2.}
pDockQ2 is a confidence score designed to predict interface quality by combining interface pLDDT with interface PAE \citep{zhu2023evaluation}. These features are first combined into
\[
x =
\frac{\mathrm{mean\ pLDDT}_{\mathrm{interface}}}
{1+(\mathrm{iPAE}/d_0)^2},
\]
and then mapped to predicted DockQ through a fitted sigmoid,
\[
\mathrm{pDockQ2}
=
\frac{L}{1+\exp[-k(x-x_0)]}+b.
\]
We use the published parameterization unless otherwise stated. Unlike ipTM, pDockQ2 was explicitly fitted against continuous DockQ, although it was not fitted separately to the individual DockQ  thresholds $\geq 0.23$, $0.49$, or $0.80$.

\paragraph{ipSAE and ipSAEmax.}
The interface predicted Score from Aligned Errors (ipSAE) is a PAE-based interface score proposed to reduce the sensitivity of ipTM to long non-interacting or poorly predicted regions \citep{dunbrack2025ipsae}. Conceptually, it applies the same TM-score-like transformation to inter-chain PAE, but restricts the calculation to residue pairs with sufficiently low PAE and determines the length normalization $d_0$ from the size of this locally confident region rather than from the full chains. For a directional chain pair $A\rightarrow B$,
\[
\mathrm{ipSAE}(A\rightarrow B)
=
\max_{i\in A}
\operatorname*{mean}_{\substack{j\in B\\
\mathrm{PAE}_{ij}<c}}
\frac{1}{
1+\left(\mathrm{PAE}_{ij}/d_0\right)^2
},
\]
where $c$ is the PAE cutoff and $d_0$ depends on the number of residues satisfying the cutoff. We use the same score implementation and cutoffs as in the corresponding source benchmarks. Because ipSAE is directional, the AFDB benchmark reports
\[
\mathrm{ipSAEmax}
=
\max\{\mathrm{ipSAE}(A\rightarrow B),
      \mathrm{ipSAE}(B\rightarrow A)\},
\]
yielding one score per complex. The analogous ``max'' convention is used for pDockQ2max in the AFDB analyses.

\section{Calibration protocol and evaluation metrics}
\label{app:calibration_details}

\paragraph{Isotonic calibration.}
For each DockQ criterion $t$, we define the binary outcome
\[
y_i^{(t)} = \mathbf{1}[\mathrm{DockQ}_i \geq t]
\]
and fit a non-decreasing function $g_t$ mapping the confidence score $s_i$ to this
outcome:
\[
g_t
= \operatorname*{arg\,min}_{g \in \mathcal{G}}
\sum_i \bigl( y_i^{(t)} - g(s_i) \bigr)^2 ,
\]
where $\mathcal{G}$ is the set of non-decreasing functions on the score domain.
A separate mapping is therefore fitted for each DockQ threshold $t\in\{0.23,0.49,0.80\}$.

For Ab--Ag, the mapping is fitted on the training portion of the temporal split and applied unchanged to the held-out temporal test set. For the within-domain general multimer evaluation, targets are partitioned into five folds. For each fold, the isotonic mapping is fitted using the other four folds and applied to the held-out fold. The resulting heldout probabilities are then concatenated to obtain the out-of-fold evaluation. Thus, no target is evaluated using a calibrator fitted on that target. Because each fitted isotonic mapping is monotone, it does not reverse the score ordering within the data to which that mapping is applied, although it can introduce ties.

\paragraph{AFDB transfer and target-domain recalibration.}
For the frozen transfer experiment, the isotonic ipTM mapping fitted on the general multimers dataset is directly applied to AFDB without refitting. For local AFDB recalibration, new isotonic mappings are fitted directly on stratified subsets of the 304 labeled AFDB complexes. Calibration-set sizes increase from 5 to 90 examples, and each fitted mapping is evaluated on the remaining held-out complexes. Each calibration size is repeated over 100 stratified splits. The same partitions are used for ipTM and ipSAEmax so that the two confidence signals are evaluated on identical calibration and hold-out sets.

\paragraph{Brier score.}
For binary outcomes $y_i^{(t)}$ and predicted probabilities $\hat p_i^{(t)}$, the Brier score is
\[
\mathrm{Brier}
=
\frac{1}{N}
\sum_{i=1}^{N}
\left(
\hat p_i^{(t)}-y_i^{(t)}
\right)^2.
\]
Lower values indicate more accurate probability estimates.

\paragraph{Expected calibration error.}
Predictions are partitioned into $B=10$ equal-width bins over $[0,1]$ according to
$\hat{p}_i^{(t)}$. For a non-empty bin $B_b$, the observed event frequency and the
mean predicted probability are
\[
\operatorname{acc}(B_b) = \frac{1}{|B_b|}\sum_{i \in B_b} y_i^{(t)},
\quad
\operatorname{conf}(B_b) = \frac{1}{|B_b|}\sum_{i \in B_b} \hat{p}_i^{(t)} .
\]
The expected calibration error is their weighted absolute difference,
\[
\mathrm{ECE}
= \sum_{b} \frac{|B_b|}{N}
\bigl| \operatorname{acc}(B_b) - \operatorname{conf}(B_b) \bigr| ,
\]
summed over non-empty bins, where $N$ is the number of evaluated predictions.
Lower ECE indicates closer agreement between predicted probabilities and observed
event frequencies.

%\section{dump figures}
\section{Calibration with Limited Ab-Ag Data}
\label{app:abag_recalibration}

\begin{figure}[!htbp]
\centering
\includegraphics[width=1.0\textwidth]{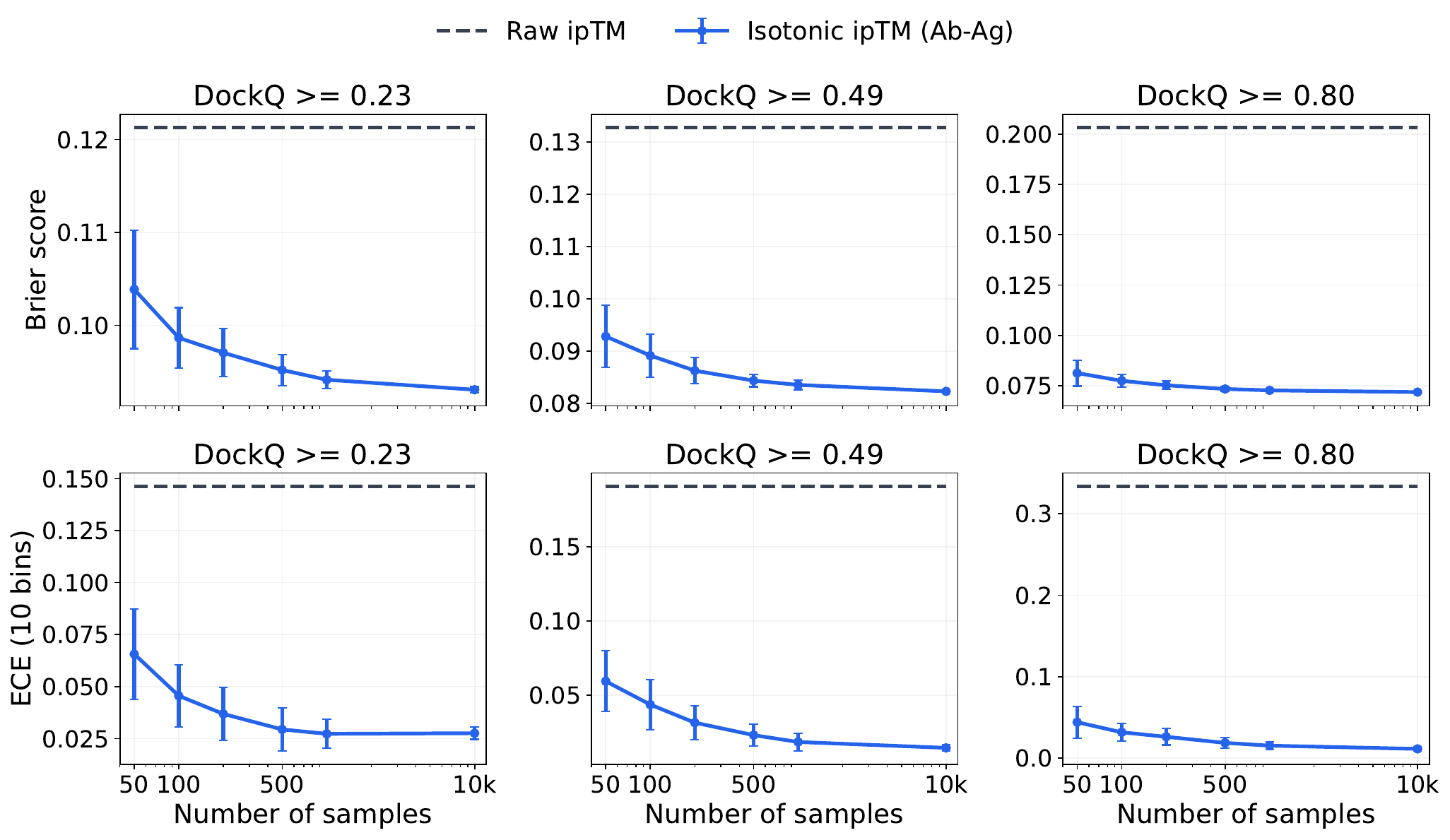}
\caption{Calibration on Ab-Ag. Brier score (top) and ECE (bottom) are shown for DockQ thresholds of 0.23, 0.49, and 0.80 as a function of the number of labeled Ab-Ag examples used for isotonic calibration. Curves show mean $\pm$ standard deviation over 100 stratified splits.}
\label{fig:abag_recalibration}
\end{figure}

We repeat local calibration on progressively larger fractions of the Ab-Ag data, as described in Sec. \ref{sec:local_calibration}. Figure~\ref{fig:abag_recalibration} shows the resulting Brier score and ECE. The improvement occurs primarily at small calibration fractions and largely saturates as additional labeled data are added.

% \begin{figure}[h]
% \centering
% \includegraphics[width=1.0\textwidth]{F1_raw_iso_largefont_v2.pdf}
% \caption{Reliability of raw and isotonically calibrated ipTM on Ab--Ag (top) and General datasets (bottom). The dashed diagonal denotes perfect calibration. Isotonic regression changes the numerical interpretation of ipTM into an event-specific estimate of $P(\mathrm{DockQ}\geq t)$, better aligning with the fraction of positives observed in the dataset.}
% \label{fig:naive_reliability}
% \end{figure}

\section{Compact Benchmark Comparisons}
\label{app:compact_tables}

\subsection{Ab--Ag baseline comparison}

Table~\ref{tab:abag_baselines} summarizes the main Ab--Ag comparison. The Brier and ECE columns quantify probability quality, whereas the AUROC column quantifies discrimination. pDockQ2 provides the strongest discrimination across the three DockQ events, while isotonic ipTM gives the lowest ECE at every threshold and the best or essentially tied Brier score.

\begin{table}[!htbp]
\caption{Ab--Ag baseline comparison on the held-out temporal test set. Lower Brier score and ECE indicate better probability estimates; higher AUROC indicates better discrimination.}
\label{tab:abag_baselines}
\centering
\small
\begin{tabular}{llccc}
\toprule
DockQ & Method & Brier$\downarrow$ & ECE$\downarrow$ & AUROC$\uparrow$ \\
\midrule
$\geq0.23$ & Raw ipTM & .1213 & .1462 & .927 \\
            %& Ranking conf. & .1478 & .1944 & .926 \\
            & pDockQ2 & .1583 & .1929 & \textbf{.943} \\
            & pDockQ2 refit & .1006 & .0969 & \textbf{.943} \\
            & ipSAEmax & .2710 & .2603 & .754 \\
            & Iso. ipTM & \textbf{.0929} & \textbf{.0282} & .926 \\
\midrule
$\geq0.49$ & Raw ipTM & .1327 & .1907 & .938 \\
           % & Ranking conf. & .1661 & .2563 & .936 \\
            & pDockQ2 & .1147 & .1260 & \textbf{.949} \\
            & pDockQ2 refit & \textbf{.0814} & .0725 & \textbf{.949} \\
            & ipSAEmax & .2886 & .3013 & .752 \\
            & Iso. ipTM & .0821 & \textbf{.0143} & .938 \\
\midrule
$\geq0.80$ & Raw ipTM & .2032 & .3336 & .932 \\
          %  & Ranking conf. & .2461 & .3992 & .928 \\
            & pDockQ2 & .0730 & .0284 & \textbf{.933} \\
            & pDockQ2 refit & .0971 & .1306 & \textbf{.933} \\
            & ipSAEmax & .3411 & .4205 & .802 \\
            & Iso. ipTM & \textbf{.0716} & \textbf{.0133} & \textbf{.933} \\
\bottomrule
\end{tabular}
\end{table}

\subsection{General baseline comparison}

General shows the same separation between discrimination and probability calibration (Table~\ref{tab:general_baselines}). The AUROC column shows that raw ipTM provides the strongest discrimination at all three DockQ thresholds, whereas the Brier and ECE columns show that isotonic ipTM provides the best probability estimates.

\begin{table}[!htbp]
\caption{General benchmark baseline comparison using target-grouped five-fold out-of-fold evaluation. Lower Brier score and ECE indicate better probability estimates; higher AUROC indicates better discrimination.}
\label{tab:general_baselines}
\centering
\small
\begin{tabular}{llccc}
\toprule
DockQ & Method & Brier$\downarrow$ & ECE$\downarrow$ & AUROC$\uparrow$ \\
\midrule
$\geq0.23$ & Raw ipTM & .1379 & .0774 & \textbf{.891} \\
           % & Ranking conf. & .1423 & .0864 & .889 \\
            & pDockQ2 & .2148 & .1990 & .827 \\
            & Iso. ipTM & \textbf{.1307} & \textbf{.0195} & .884 \\
\midrule
$\geq0.49$ & Raw ipTM & .1728 & .1957 & \textbf{.913} \\
           % & Ranking conf. & .1806 & .2080 & .912 \\
            & pDockQ2 & .1645 & .0506 & .843 \\
            & Iso. ipTM & \textbf{.1186} & \textbf{.0086} & .907 \\
\midrule
$\geq0.80$ & Raw ipTM & .3224 & .4428 & \textbf{.876} \\
           % & Ranking conf. & .3313 & .4557 & .875 \\
            & pDockQ2 & .1781 & .2044 & .838 \\
            & Iso. ipTM & \textbf{.1055} & \textbf{.0118} & .869 \\
\bottomrule
\end{tabular}
\end{table}

\subsection{Frozen General-to-AFDB transfer}

Table~\ref{tab:afdb} summarizes the frozen General-to-AFDB transfer experiment. The Brier and ECE columns show that the General-trained probability mapping does not retain a consistent calibration advantage on AFDB, whereas the AUROC column shows that the confidence signals remain informative for discrimination. The AUROC column also shows that ipSAEmax provides the strongest discrimination on AFDB across the three DockQ events.

\begin{table}[!htbp]
\caption{Frozen transfer to AFDB on 304 complexes with finite DockQ. The General-trained isotonic ipTM mapping is applied without refitting.}
\label{tab:afdb}
\centering
\small
\begin{tabular}{llccc}
\toprule
DockQ & Method & AUROC$\uparrow$ & Brier$\downarrow$ & ECE$\downarrow$ \\
\midrule
$\geq0.23$ & Raw ipTM & .886 & .142 & .102 \\
            & General$\rightarrow$AFDB Iso. & .884 & .156 & .116 \\
            & pDockQ2max & .887 & .281 & .302 \\
            & ipSAEmax & \textbf{.911} & \textbf{.133} & .122 \\
\midrule
$\geq0.49$ & Raw ipTM & .879 & .146 & .092 \\
            & General$\rightarrow$AFDB Iso. & .878 & .187 & .163 \\
            & pDockQ2max & .866 & .227 & .218 \\
            & ipSAEmax & \textbf{.894} & \textbf{.133} & \textbf{.082} \\
\midrule
$\geq0.80$ & Raw ipTM & .865 & .212 & .309 \\
            & General$\rightarrow$AFDB Iso. & .857 & .112 & .081 \\
            & pDockQ2max & .876 & \textbf{.100} & \textbf{.063} \\
            & ipSAEmax & \textbf{.892} & .178 & .212 \\
\bottomrule
\end{tabular}
\end{table}

\FloatBarrier

\FloatBarrier

\section{pDockQ2 Refit}
\label{app:pdockq2_refit}

To distinguish event-specific calibration from simply adapting an existing continuous interface-quality score to the Ab--Ag domain, we refit the published pDockQ2 on the Ab-Ag dataset. The training data contain 269,105 rows across 7045 targets, of which 248,670 have valid values for the continuous-DockQ fit. The regression target is continuous DockQ rather than any of the three threshold-specific binary events.

The pDockQ2 functional form is
\[
\mathrm{pDockQ2}
=
\frac{L}{1+\exp[-k(x-x_0)]}+b,
\qquad
x=
\frac{\mathrm{mean\_pLDDT}_{\mathrm{int}}}
{1+(\mathrm{iPAE}/d_0)^2},
\]
with $d_0=10$ fixed. Parameters $L$, $x_0$, $k$, and $b$ are fitted using nonlinear least squares with the published values as initialization, $L=1.31$, $x_0=84.733$, $k=0.075$, and $b=0.005$. Optimization uses the bounds $L\in[0,2]$, $x_0\in[0,200]$, $k\in[0,1]$, and $b\in[-1,1]$, and the final output is clipped to $[0,1]$ only after fitting.

The resulting parameters are
\[
L=2.00000000,\qquad
k=0.01947584,\qquad
x_0=40.21941805,\qquad
b=-0.65947117.
\]

The fitted value of $L$ reaches the specified upper optimization bound and is reported without further interpretation. No temporal-test observations are used during fitting. We refer to this model as \emph{refitted pDockQ2}, rather than calibrated pDockQ2, because it remains a single continuous prediction of interface quality rather than a separate estimate of $P(\mathrm{DockQ}\geq t)$ for each structural-quality event.

\section{Ranking Preservation}
\label{sec:ranking_preservation}

Isotonic regression is monotone and therefore cannot introduce strict reversals in the ordering implied by its input score. Its piecewise-constant mapping can introduce ties, which can produce small differences in rank-based metrics or alter selections exactly at an operating-point boundary. Small AUROC differences after calibration should therefore be attributed to ties rather than to new ranking information introduced by the calibrator.

\FloatBarrier

% If required by the workshop/submission system:
% \input{checklist.tex}

\end{document}